\documentclass[aps,pra,twocolumn,groupedaddress,showpacs,amsmath,amssymb]{revtex4}

\usepackage{graphicx}
\usepackage{amsmath}
\usepackage{bm}

\def\6#1{{\underline{#1}}}
\def\m6#1{{\underline{#1}\,}}

\newdimen\Tdim
\def\ispan{{\setbox0=\hbox{i}%
\Tdim\ht0\advance\Tdim\dp0\rule[-\dp0]{0pt}{\Tdim}}}
\def\jspan{{\setbox0=\hbox{j}%
\Tdim\ht0\advance\Tdim\dp0\rule[-\dp0]{0pt}{\Tdim}}}
\def\Tspan#1{{\setbox0=\hbox{#1}%
\Tdim\ht0\advance\Tdim\dp0\advance\Tdim.55ex\rule[-\dp0]{0pt}{\Tdim}\box0}}

\def\be{\begin{eqnarray}}
\def\ben{\begin{eqnarray*}}
\def\ee{\end{eqnarray}}
\def\een{\end{eqnarray*}}

\def\p{\partial}

\def\=:{=\hspace{-.7em}\raisebox{1.1ex}{.}\hspace{.1em}\raisebox{-0.2ex}{.} }

\newcommand {\1}[1]{\frac{1}{#1}}

\newcommand {\beq}{\begin{eqnarray}}
\newcommand {\eeq}{\end{eqnarray}}
\newcommand {\non}{\nonumber\\}

\begin{document}


\title{Interaction of half-quantized vortices in two-component Bose-Einstein condensates}


\author{Minoru Eto$^{1}$}
\author{Kenichi Kasamatsu$^{2}$}
\author{Muneto Nitta$^{3}$}
\author{Hiromitsu Takeuchi$^{4}$}
\author{Makoto Tsubota$^{4}$}
\affiliation{
$^1$Theoretical Physics Laboratory, RIKEN, Saitama 351-0198, Japan \\
$^2$Department of Physics, Kinki University, Higashi-Osaka, 577-8502, Japan \\
$^3$Department of Physics, and Research and Education Center for Natural 
Sciences, Keio University, Hiyoshi 4-1-1, Yokohama, Kanagawa 223-8521, Japan \\
$^4$Department of Physics, Osaka City University, Sumiyoshi-Ku, Osaka 558-8585, Japan}


\date{\today}

\begin{abstract}
We study the asymptotic interaction 
between two half-quantized vortices 
in two-component Bose-Einstein condensates. 
When two vortices in different components 
are placed at distance $2R$,  
the leading order of the force between them is found to be 
$(\log R/\xi-1/2)/R^3$, in contrast to $1/R$ between 
vortices placed in the same component.
We derive it analytically using 
the Abrikosov ansatz and the profile functions 
of the vortices, confirmed numerically 
with the Gross-Pitaevskii model.
We also find that the short-range cutoff of 
the inter-vortex potential 
linearly depends on the healing length.
\end{abstract}


\maketitle

\section{Introduction}
Multicomponent condensations appear in 
many systems in condensed matter physics and QCD, 
from multi-component or spinor Bose-Einstein condensates (BECs), 
superfluid $^3$He, multi-gap superconductors to 
chiral phase transition or color superconductors in QCD 
at high temperature and/or high density.
Especially, multicomponent and spinor BECs admit 
a rich variety of topological excitations: 
Domain walls \cite{Tim}, 
Abelian \cite{Kasamatsu2} and 
non-Abelian \cite{Semenoff:2006vv} vortices, 
monopoles \cite{monopoles}, 2D Skyrmions \cite{2D-Skyrmions}, 
3D Skyrmions \cite{3D-Skyrmions,3D-Skyrmions2}, 
vortons \cite{Metlitski:2003gj}, knots \cite{Kawaguchi}, 
and D-brane solitons \cite{Kasamatsu:2010aq}. 
See \cite{Kasamatsu2,Ueda-Kawaguchi,Kawaguchi:2010mu} as a review.
Among these topological excitations, quantized vortices in multicomponent 
BECs are the most important subject, because they are closely 
related to the problems not only in other condensed matter systems 
such as superconductors, superfluids, magnetism, and liquid crystal, 
but also in electro-weak theory \cite{Achucarro:1999it}, QCD 
and grand unified theories in high energy physics, neutron stars   
and cosmic strings in cosmology \cite{Hindmarsh:1994re,Vilenkin-Shellard}. 

Interactions between quantized vortices are important information 
to determine the equilibrium configuration and dynamics 
of many vortices. It is known that, in a single-component BEC, the asymptotic 
interaction energy per unit length of two parallel vortex lines separated by a distance 
$R$ is proportional to $\log (L/R)$, where $L$ is the size of the system \cite{Pethickbook}.  
Thus, the inter-vortex force has $1/R$ dependence.
Vortices in a BEC resemble with global vortices in relativistic field theories 
\cite{Vilenkin:1982ks}--\cite{Davis:1989gn}.
A relation between them was studied in \cite{Davis:1989gn} 
where it was suggested that spinning global vortices on a Lorentz violating 
background behave as superfluid vortices.  
Global vortices are regarded as global cosmic strings or axion strings 
in cosmology and the inter-vortex force between two global vortices 
was shown to be $1/R$ \cite{Shellard:1987bv}, coinciding with the one in vortices 
in a scalar BEC, scalar superfluid, and the XY model. 
Global vortices also appear in QCD; in chiral phase transition of QCD 
at high temperature or high density \cite{Balachandran:2002je,Nakano:2007dq}
or color superconductor of extremely high density QCD
\cite{Balachandran:2005ev,Nakano:2007dr}. Inter-vortex force at large distance $R$
was derived analytically at the leading order as $1/R$ for color 
superconductor \cite{Nakano:2007dr}, and $\cos \alpha /R$ with 
a relative orientation $\alpha$ of two vortices in the internal space 
for chiral phase transition \cite{Nakano:2007dq};  see \cite{Nakano:2008dc} as a review. 

However, the analytic formula of the vortex-vortex interactions in multicomponent 
BECs are still missing. Two-component BECs are the simplest example of the
multicomponent condensates and have also attracted much
interest to study the novel phenomena not found in a 
single component BEC. 
Recent experiments provide a good ground of study on the vortex-vortex interaction in two-component BECs
by tuning s-wave scattering length via 
a Feshbach resonance \cite{Thalhammer,Papp,Tojo}. 
The minimally quantized vortex in two-component BECs 
has the winding number one half of a singly-quantized vortex  
in scalar BECs, and thus is often called a half-quantized vortex.  
Its mass circulation is fractionally quantized 
when mass densities of two condensates are different. 
Such a quantized vortex in two-component BECs has a composite structure, 
where a vortex core in one component is filled by the density of the 
other component. This vortex structure was created experimentally 
through coherent interconversion between two components \cite{Matthews}. 
Interactions between the vortices in the different components are nontrivial 
because the two components interact only through the density, 
so that 
the vortex winding around one component 
does not directly experience the circulation of the 
other vortex winding around the other component. 
This fact results in an indirect interaction, where the filling component 
of each vortex core is affected by the circulation created by the vortex 
in the {\it same} component, dragging the vortex in which it is filled. 
Although the interactive dynamics of two vortices in two-component BECs 
was studied numerically by \"{O}hberg and Santos \cite{OS2002}, 
the analytical form of the interaction force was not discussed. 

In this paper, we consider the asymptotic interaction between two vortices 
in two-component BECs. We consider the vortex-vortex interaction for two cases 
(i) two vortices are placed in the different components 
and (ii) those in the same component. For the case (i), the leading 
order of the inter-vortex force between them at distance $2R$ is found to be $(\log R/\xi - 1/2)/R^3$
with the short-range cutoff $\xi$, 
in contrast to the one $1/R$ for the case (ii) and vortices in a single-component BEC.  
We derive it analytically using the Abrikosov ansatz and the asymptotic profile functions 
of the vortices. We then confirm it numerically. 
We also find that the short-range cutoff $\xi$ of 
the inter-vortex potential linearly depends 
on the healing length.

This paper is organized as follows. Sec.~II is devoted for deriving the analytic
form of the asymptotic inter-vortex force. In Sec.~III we confirm the analytic results 
obtained in Sec.~II by numerical calculations of the Gross-Pitaevskii equation.
Summary and discussions are in Sec.~IV. 
In Appendix, we describe some details of the calculation of integrals 
in Sec.~II C.

\section{Static Inter-vortex Forces}

\subsection{The model}

We start with an energy functional for two-component BEC system
\beq
&& E(\Psi_1,\Psi_2) = K(\Psi_1,\Psi_2)+V(\Psi_1,\Psi_2),\\
&& K = \int d^3x \sum_{i=1,2}\left(
- \frac{\hbar^2}{2m_i} \Psi^*_i \nabla^2 \Psi_i 
\right),\\
&& V = \int d^3x \left[\sum_{i=1,2}
\frac{g_{i}}{2} |\Psi_i|^4 
+ g_{12} |\Psi_1|^2|\Psi_2|^2 \right],
\label{eq:GP_energy}
\eeq
where $\Psi_i$ is a condensate wave function 
of the $i$-th component ($i=1,2$) with mass $m_i$.
The coupling constants $g_1,g_2$ and $g_{12}$ stand for 
the atom-atom interactions; the $\Psi_1$ and $\Psi_2$ components 
repel or attract for $g_{12}>0$ or $g_{12} < 0$, respectively.
The coupled Gross-Pitaevskii (GP) equations are obtained by the
variational principle $i\hbar \p_t\Psi_i = \delta E/\delta \Psi_i^*$ as
\beq
i\hbar \p_t \Psi_1 &=& \left( -\frac{\hbar^2 \nabla^2}{2m_1} + g_1 |\Psi_1|^2 + g_{12}|\Psi_2|^2\right)\Psi_1,
\label{eq:GP1}\\
i\hbar \p_t \Psi_2 &=& \left( -\frac{\hbar^2 \nabla^2}{2m_2} + g_2 |\Psi_2|^2 + g_{12}|\Psi_1|^2\right)\Psi_2,
\label{eq:GP2}
\eeq
The stationary coupled GP equation is given 
by considering a time dependence $\Psi_i({\bf x},t) 
= e^{-i\mu_i t/\hbar} \Psi_i({\bf x})$ with
the chemical potential $\mu_i$
\beq
\left( -\frac{\hbar^2 \nabla^2}{2m_1} - \mu_1 + g_1 |\Psi_1({\bf x})|^2 + g_{12}|\Psi_2({\bf x})|^2\right)\Psi_1({\bf x}) &=& 0,
\label{eq:GP3}\nonumber\\ \\
\left( -\frac{\hbar^2 \nabla^2}{2m_2} - \mu_2 + g_2 |\Psi_2({\bf x})|^2 + g_{12}|\Psi_1({\bf x})|^2\right)\Psi_2({\bf x}) &=& 0.
\label{eq:GP4}\nonumber\\
\eeq

The potential energy $V$ with the quadratic terms $-\mu_1|\Psi_1|^2 - \mu_2 |\Psi_2|^2$
induced by the chemical potential
is a quadratic function of 
$X \equiv |\Psi_1|^2\ge0$ and $Y \equiv |\Psi_2|^2\ge0$,
\beq
V(X,Y) = \frac{g_1}{2}X^2 + \frac{g_2}{2}Y^2 + g_{12}XY - \mu_1 X - \mu_2 Y.
\eeq
Let $g_1,g_2$ be positive, then
the potential $V$ has a minimum when 
\begin{eqnarray}
\Delta \equiv V_{XX}V_{YY}- V_{XY}^2 = g_1g_2 - g_{12}^2 > 0, \nonumber \\ 
\mu_1g_2-\mu_2g_{12} \ge 0, \quad  \mu_2g_1-\mu_1g_{12} \ge 0. 
\label{inequility}
\end{eqnarray}
The amplitudes of the ground state are then given by
\beq
|\Psi_1| = \sqrt{\frac{\mu_1 g_2 - \mu_2 g_{12}}{g_1g_2-g_{12}^2}} \equiv v_1, \nonumber \\
|\Psi_2| = \sqrt{\frac{\mu_2 g_1 - \mu_1 g_{12}}{g_1g_2-g_{12}^2}} \equiv v_2.
\label{eq:v}
\eeq

In the following, we consider the situation in which the above inequalities are satisfied. 
Since there are two condensates, 
two $U(1)$ symmetries are spontaneously broken.
Accordingly the order parameter space is 
\beq
 T^2 \simeq U(1)_1 \times U(1)_2 \simeq 
{U(1)_{\rm mass} \times U(1)_{\rm spin} \over {\mathbb Z}_2}.
 \label{eq:OPS}
\eeq 
Here each $U(1)_i$ ($i=1,2$) corresponds
to the phase rotation of $\Psi_1$ or $\Psi_2$, while
$U(1)_{\rm mass}$ and $U(1)_{\rm spin}$ 
correspond to the overall and relative phase rotations, defined by 
\beq
 U(1)_{\rm mass}:&& \quad \Psi_1 \to \Psi_1 e^{i \alpha}, \quad 
 \Psi_2 \to \Psi_2 e^{i \alpha}, \nonumber \\
 U(1)_{\rm spin}:&& \quad \Psi_1 \to \Psi_1 e^{i \beta}, \quad 
 \Psi_2 \to \Psi_2 e^{-i \beta} ,\label{eq:mass-spin}
\eeq
whose currents are mass and pseudo-spin currents, respectively.
Both the condensates $\Psi_1,\Psi_2$ are unchanged under 
the ${\mathbb Z}_2$ action ($\alpha = \beta = \pi$)
inside $U(1)_{\rm mass} \times U(1)_{\rm spin}$ 
in Eq.~(\ref{eq:mass-spin}), 
and therefore this ${\mathbb Z}_2$ has to be removed, 
as the denominator of Eq.~(\ref{eq:OPS}).

In what follows, we call the phase cycles for $\Psi_1$ and $\Psi_2$ 
the $(1,0)$- and ${(0,1)}$-cycles, respectively.

\subsection{Vortex configuration}

Since the first homotopy group of order parameter space is 
\beq 
 \pi_1(T^2) = {\mathbb Z} \oplus {\mathbb Z},
\eeq 
it allows two kinds of winding numbers.
We refer a vortex winding around $(1,0)$[${(0,1)}$]-cycle once as 
a $(1,0)$-vortex [$(0,1)$-vortex], 
which is the most fundamental vortex. 
When one travels around a $(1,0)$[${(0,1)}$]-vortex, 
the phase of $\Psi_1$($\Psi_2$) rotates by $2\pi$ with 
the phase of the other component kept constant. 
On the other hand, 
in terms of $U(1)_{\rm mass}$ and $U(1)_{\rm spin}$ in 
Eq.~(\ref{eq:mass-spin}), 
$U(1)_{\rm mass}$ is rotated by $\pi$ and 
$U(1)_{\rm spin}$ is rotated by $+ \pi$ ($- \pi$) 
with circling around a $(1,0)$[${(0,1)}$]-vortex. 
Since they have a half winding of $U(1)_{\rm mass}$, 
they are often called {\it half-quantized} vortices.

Vortices winding around both components by $2\pi$ 
are denoted by $(1,1)$ and 
have unit winding in $U(1)_{\rm mass}$.
They are called integer vortices,
if the core is not separated into 
$(1,0)$ and $(0,1)$ vortices.
More generally we refer 
a configuration
which winds $(1,0)$-cycle $m$ times and ${(0,1)}$-cycle $n$ times
as an $(m,n)$-vortex, whose wave function is denoted as 
$\Psi_i^{(m,n)}$ for $i$-th component.

The vortex configuration can be obtained by solving Eqs.~(\ref{eq:GP3}) and (\ref{eq:GP4}).
Let us make an ansatz for an axially symmetric $(1,0)$-vortex configuration 
\beq
\Psi_1^{(1,0)} = v_1\ e^{i\theta} f_{(1,0)}(r),\quad 
\Psi_2^{(1,0)} = v_2\  h_{(1,0)}(r),
\label{eq:ansatz}
\eeq
where $r$ and $\theta$ are the polar coordinates.
\begin{widetext}
The profile functions $f_{(1,0)}$ and $h_{(1,0)}$ 
are determined by substituting 
(\ref{eq:ansatz}) into (\ref{eq:GP3}) and (\ref{eq:GP4}), as
\beq
- \frac{\hbar^2}{2m_1} \left(f_{(1,0)}'' + \frac{f_{(1,0)}'}{r} - \frac{f_{(1,0)}}{r^2} \right)
+ \frac{\mu_1g_1g_2 (f_{(1,0)}^2 - 1)
- \mu_1 g_{12}^2 (h_{(1,0)}^2-1) 
- \mu_2 g_1g_{12}(f_{(1,0)}^2-h_{(1,0)}^2)}{g_1g_2-g_{12}^2} f_{(1,0)} = 0,
\label{eq:GP_axial_1}\\
- \frac{\hbar^2}{2m_2} \left(h_{(1,0)}'' + \frac{h_{(1,0)}'}{r} \right)
+ \frac{\mu_2g_1g_2 (h_{(1,0)}^2 - 1)
- \mu_2 g_{12}^2 (f_{(1,0)}^2-1) 
- \mu_1 g_2g_{12}(h_{(1,0)}^2-f_{(1,0)}^2)}{g_1g_2-g_{12}^2}h_{(1,0)} = 0, 
\label{eq:GP_axial_2}
\eeq
\end{widetext}
with the prime denoting a differentiation with respect to $r$. 
We solve these equations with the boundary conditions
\beq
(f_{(1,0)},h_{(1,0)}) &\to& (1,1) \qquad \text{as}\quad r \to \infty,\\
(f_{(1,0)},h'_{(1,0)}) &\to& (0,0) \qquad \text{as}\quad r \to 0.
\eeq
From these equations, asymptotic behaviors of the profile functions 
$f_{(1,0)}$ and $h_{(1,0)}$ at large distance 
can be obtained as 
\beq
f_{(1,0)} (r) &=& 1 - \frac{1}{m_1\eta_1^+ r^2} + {\cal O}(r^{-4}),
\label{eq:gp_asym1}\\
h_{(1,0)} (r) &=& 1 + \frac{1}{m_1 \eta_1^- r^2} + {\cal O}(r^{-4}),
\label{eq:gp_asym2}
\eeq
where we have introduced the effective mass parameters
\beq
\eta_1^+ \equiv \frac{4(\mu_1 g_2 - \mu_2 g_{12})}{g_2\hbar^2},\quad
\eta_1^- \equiv \frac{4(\mu_2 g_{1} - \mu_1 g_{12})}{g_{12}\hbar^2}.
\eeq
The stability condition Eq.~(\ref{inequility}) of the ground state ensures 
that $\eta_1^+ >0$, while $\eta_1^-$ changes its sign with $g_{12}$.

Similarly, we make an ansatz for the $(0,1)$-vortex 
\beq
\Psi_1^{(0,1)} =v_1 h_{(0,1)}(r),\quad \Psi_2^{(0,1)} = v_2 e^{i\theta} f_{(0,1)}(r).
\label{eq:ansatz_b}
\eeq
The equations for $f_{(0,1)},h_{(0,1)}$ can be obtained by just replacing the indices
as $1 \leftrightarrow 2$ and ${(1,0)} \leftrightarrow {(0,1)}$ in Eqs.~(\ref{eq:GP_axial_1})
and (\ref{eq:GP_axial_2}).
Then the asymptotic behaviors are
\beq
f_{(0,1)} (r) &=& 1 - \frac{1}{m_2\eta_2^+r^2} + {\cal O}(r^{-4}),
\label{eq:gp_asym3}\\
h_{(0,1)} (r) &=& 1 + \frac{1}{m_2\eta_2^-r^2} + {\cal O}(r^{-4}),
\label{eq:gp_asym4}
\eeq
with 
\beq
\eta_2^+ \equiv \frac{4(\mu_2 g_1 - \mu_1 g_{12})}{g_1\hbar^2},\quad
\eta_2^- \equiv \frac{4(\mu_1 g_{2} - \mu_2 g_{12})}{g_{12}\hbar^2}.
\eeq
Again, $\eta_2^+$ is always positive while sign of $\eta_2^-$ depends on $g_{12}$.

As vortices in a scalar BEC, 
the tension (energy per unit length) of 
$(1,0)$ and $(0,1)$ vortices 
logarithmically diverges as 
\beq
T_{(1,0)} \simeq 
\frac{\pi \hbar^2 v_1^2}{m_1} \log \frac{L}{\xi},\qquad
T_{(0,1)} \simeq 
\frac{\pi \hbar^2 v_2^2}{m_2} \log \frac{L}{\xi},
\eeq
respectively, 
with $L$ and $\xi$ being a long and short  distance cutoff, respectively.
This divergent behavior comes from the kinetic term in the GP energy 
functional Eq.~(\ref{eq:GP_energy}).
\begin{figure}[ht]
\begin{center}
\includegraphics[width=7cm]{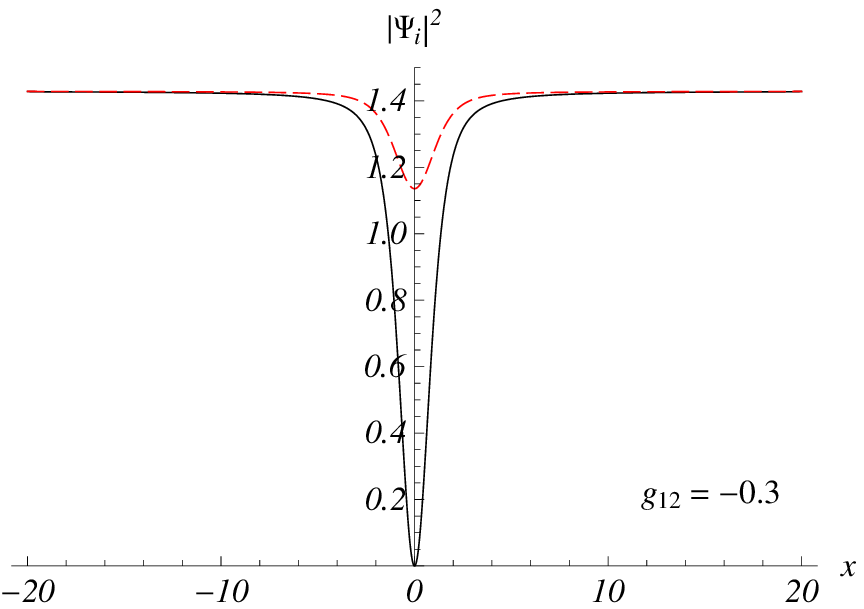}\\\ (a) \\\
\includegraphics[width=7cm]{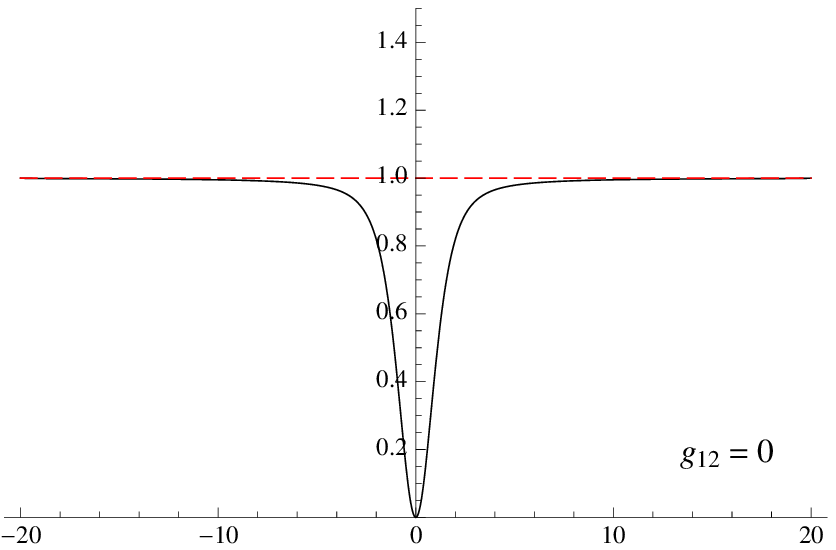}\\\ (b) \\\ 
\includegraphics[width=7cm]{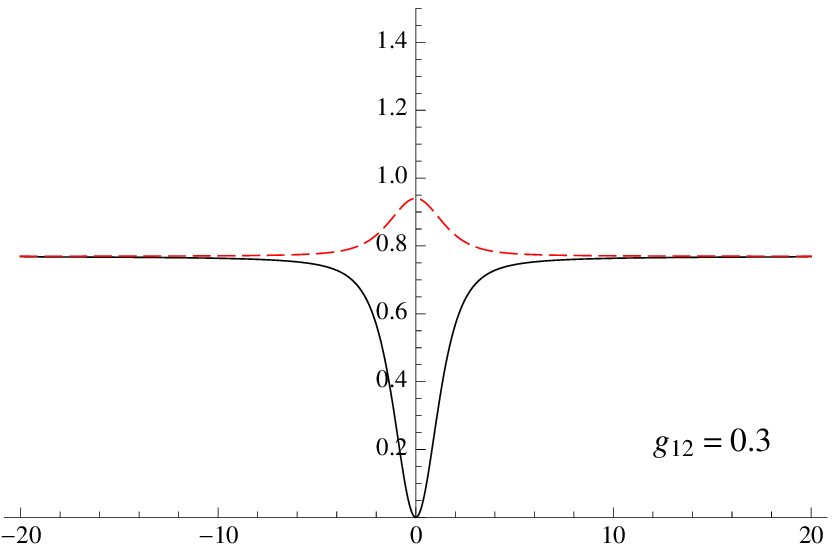}\\\ (c)
\caption{ Single vortex configurations ($|\Psi_1|^2$ (solid line) and $|\Psi_2|^2$ (broken line)) on a cross section.  
The field $\Psi_1$ winds once
so that $|\Psi_1|$ goes to zero at the vortex center 
while $\Psi_2$ does not touch zero anywhere but it can
have nonzero amplitude at the vortex center.
The parameters are $(\hbar,g_1,g_2,\mu_1,\mu_2,m_1,m_2)=(1,1,1,1,1,1,1)$ 
and (a) $g_{12} = -0.3$, (b) $g_{12}=0$ and (c) $g_{12}=0.3$.}
\label{fig:config}
\end{center}
\end{figure}

Some numerical solutions of the single vortex configurations are shown in 
Fig.~\ref{fig:config}.
A universal feature of configuration is that $h_{(1,0)}$ (the profile function of unwinding field)
at the vortex center is concave for $g_{12}<0$ 
and convex for the $g_{12} > 0$ \cite{footnote1}. 
This can be understood from the atom-atom interaction $g_{12}$; 
in the presence of the vortex profile for $\Psi_1$ as a background, 
$\Psi_2$ feels the potential $g_{12}|\Psi_1|^2$
and it tends to be trapped in the vortex center for 
the repulsive interaction $g_{12}>0$ 
and to be exclusive from the vortex center for the 
attractive interaction $g_{12}<0$.

\subsection{Inter-vortex forces}

\begin{figure}[ht]
\begin{center}
\bigskip\bigskip
\includegraphics[height=4cm]{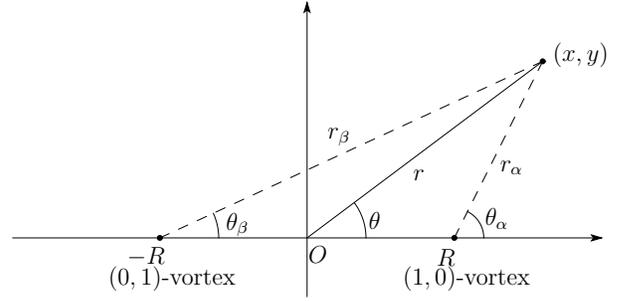}
\caption{Configuration of $(1,0)$-vortex 
and $(0,1)$-vortex.}
\label{fig:conf}
\end{center}
\end{figure}
It is expected that the interactions between $(1,0)$- and $(0,1)$-vortices are 
determined by the coupling $g_{12}$-term.
When $g_{12}$ is zero,
they are decoupled in Eqs.~(\ref{eq:GP1}) and (\ref{eq:GP2}), 
so $(1,0)$ and $(0,1)$ vortices do not interact. 
Here, we calculate the asymptotic interactions 
between well separated $(1,0)$- and $(0,1)$-vortices 
using their asymptotic profile functions obtained in 
the last subsection.
Let us place the $(1,0)$- and $(0,1)$-vortices at $(x,y)= (R,0)$ and 
$(x,y)=(-R,0)$, respectively as in Fig.~\ref{fig:conf}. 
We use the polar coordinates $(r,\theta)$ with the origin $(x,y)=(0,0)$. 
We further express $(r_{{(1,0)}},\theta_{{(1,0)}})$ and 
$(r_{{(0,1)}},\theta_{{(0,1)}})$ 
as the polar coordinates with the origins at the $(1,0)$ and $(0,1)$ 
vortex centers 
$(R,0)$ and $(-R,0)$, respectively. 
Then the following relations hold among three polar coordinates 
\beq
&&r_i^2 = \left(r\cos\theta \mp R\right)^2 + r^2 \sin^2\theta,\nonumber\\
&&\tan\theta_i = \frac{r\sin\theta}{r\cos\theta \mp R} 
\eeq
with $i={(1,0)},{(0,1)}$, the minus sign for $i={(1,0)}$ 
and the plus sign for $i={(0,1)}$.
With these coordinates, 
the $(1,0)$- and $(0,1)$-vortex configurations 
$(\Psi_1^{(1,0)},\Psi_2^{(1,0)})$ and
$(\Psi_1^{(0,1)},\Psi_2^{(0,1)})$ 
can be expressed as
\beq
&& \Psi_1^{(1,0)} = v_1
e^{i\theta_{(1,0)}} f_{(1,0)}(r_{(1,0)}),\nonumber\\
&& \Psi_2^{(1,0)} = v_2
h_{(1,0)}(r_{(1,0)}),\\
&& \Psi_1^{(0,1)} = v_1
h_{(0,1)}(r_{(0,1)}),\nonumber\\
&& \Psi_2^{(0,1)} = v_2
e^{i\theta_{(0,1)}} f_{(0,1)}(r_{(0,1)}).
\eeq

\begin{widetext}
Let us now calculate the interaction between 
$(1,0)$-vortex and $(0,1)$-vortex.
We first make the standard Abrikosov ansatz 
\beq
\Psi_1^{(1,1)}(r,\theta) &=& v_1^{-1} \Psi_1^{(1,0)}\Psi_1^{(0,1)}
\simeq v_1\left(1 - \frac{1}{m_1\eta_1^+r_{(1,0)}^2} 
+ \frac{1}{m_2\eta_2^- r_{(0,1)}^2} \right)e^{i\theta_{(1,0)}} 
+ {\cal O}(r^{-4}),
\label{eq:abrikosov1}\\
\Psi_2^{(1,1)}(r,\theta) &=& v_2^{-1} \Psi_2^{(1,0)}\Psi_2^{(0,1)} 
\simeq v_2\left(1 - \frac{1}{m_2\eta_2^+r_{(0,1)}^2} 
+ \frac{1}{m_1\eta_1^- r_{(1,0)}^2} \right)e^{i\theta_{(0,1)}}
+ {\cal O}(r^{-4}) 
\label{eq:abrikosov2}
\eeq
for the total configuration.
\end{widetext}
Then the interaction potential is obtained by 
subtracting two individual vortex energies
from the total energy as 
\beq
U_{(1,1)} = \int d^2x \left(\delta K + \delta V\right), 
\label{eq:potentia}
\eeq
where it has two contributions: one from the kinetic energy 
$\delta K = 
K(\Psi_1^{(1,1)},\Psi_2^{(1,1)})-K(\Psi_1^{(1,0)},\Psi_2^{(1,0)})-K(\Psi_1^{(0,1)},\Psi_2^{(0,1)})$
and the other from the potential energy 
$\delta V = 
V(\Psi_1^{(1,1)},\Psi_2^{(1,1)})-V(\Psi_1^{(1,0)},\Psi_2^{(1,0)})-V(\Psi_1^{(0,1)},\Psi_2^{(0,1)})+V(v_1,v_2)$.

By using the asymptotic properties given in Eqs.~(\ref{eq:gp_asym1}), (\ref{eq:gp_asym2}),
(\ref{eq:gp_asym3}), (\ref{eq:gp_asym4}), (\ref{eq:abrikosov1}) and (\ref{eq:abrikosov2}), we find
\beq
\delta K &=& 
\frac{v_1^2\hbar^2}{m_1m_2\eta_2^-r_{(0,1)}^2}(\nabla\theta_{(1,0)})^2 
\nonumber\\
&& 
+ \frac{v_2^2\hbar^2}{m_1m_2\eta_1^-r_{(1,0)}^2}(\nabla\theta_{(0,1)})^2 
+ {\cal O}(r^{-6})
\non
&=& \frac{g_{12} \hbar^4}{2m_1m_2(g_1g_2-g_{12}^2)r_{(1,0)}^2 r_{(0,1)}^2}
+ {\cal O}(r^{-6}),
\eeq
where we have used $(\nabla \theta_{(1,0)})^2 = r_{(1,0)}^{-2},\ (\nabla \theta_{(0,1)})^2 = r_{(0,1)}^{-2}$
and have taken terms up to ${\cal O}(r^{-4})$.
It is important to see that the leading terms of the order ${\cal O}(r^{-2})$ 
have been canceled out in the subtraction.
Therefore, the dominant contribution to the interaction potential is of the order ${\cal O}(r^{-4})$. 
Similarly, we find the terms of the order ${\cal O}(r^{-4})$ in the potential energy
\beq
\delta V = - \frac{g_{12} \hbar^4}{4m_1m_2(g_1g_2-g_{12}^2)r_{(1,0)}^2 r_{(0,1)}^2}
+ {\cal O}(r^{-6}). \label{intreactionpot}
\eeq
Plugging these into Eq.~(\ref{eq:potentia}), we get
\beq
U_{(1,1)}(R) &=& \frac{g_{12} \hbar^4}{4m_1m_2(g_1g_2-g_{12}^2)} 
\int d^2x\ \frac{1}{r_{(1,0)}^2r_{(0,1)}^2} \nonumber \\ 
&\simeq& \frac{g_{12} \hbar^4 \pi}{4m_1m_2(g_1g_2-g_{12}^2)} 
{\log \frac{R}{\xi} \over R^2},
\label{eq:pot}
\eeq
where $\xi$ stands for a short distance cut-off and we have used
$R \gg \xi$. The detailed calculation of Eq.~(\ref{eq:pot}) is described in Appendix. 
Here, the terms independent of $R$ have been ignored. 
The factor $1/R^2$ is a striking feature
which is absent in the scalar BEC or scalar superfluids. 
Note that the chemical potentials $\mu_1$ and $\mu_2$ do not appear in the final result (\ref{eq:pot}).
The asymptotic force between the two vortices 
is obtained by differentiating the potential 
by their distance $2R$, as $F_{(1,1)}(R) = - \frac{dU_{(1,1)}}{2dR}$,
\beq
F_{(1,1)}(R) = \frac{\pi \hbar^4 g_{12}}{4m_1m_2(g_1g_2-g_{12}^2)} \frac{1}{R^3}  
\left( \log\frac{R}{\xi} - \frac{1}{2} \right).
\label{eq:(1,1)force}
\eeq
We have found that the interaction is attractive  for $g_{12} < 0$, 
repulsive  for $g_{12} > 0$ and vanishes for $g_{12}=0$.

Note that the asymptotic interaction is independent of the sign of 
the vortex winding number $e^{\pm i \theta}$,  
because the interaction  
between the two condensates is mediated only
through their amplitudes as $g_{12}|\Psi_1|^2|\Psi_2|^2$.
In fact, the interaction potential $U_{(1,-1)}$ between
$(1,0)$ and $(0,-1)$ vortices are exactly the same as $U_{(1,1)}$.
It is easy to verify that the following relation holds
\beq
U_{(1,1)} = U_{(1,-1)} = U_{(-1,1)} = U_{(-1,-1)}.
\eeq 
This is because $\theta_{(1,0)}$ and $\theta_{(0,1)}$ are decoupled in the 
Abrikosov ansatz in Eqs.~(\ref{eq:abrikosov1}) and (\ref{eq:abrikosov2}).

The potential (\ref{eq:pot}) should be compared with 
the potential $U_{(1\pm1,0)}$ between $(1,0)$ and $(\pm 1,0)$ vortices.
To see it, we make the ordinary Abrikosov ansatz
\beq
\Psi_1^{(1\pm1,0)} \simeq v_1 e^{i(\theta_{(1,0)} \pm \theta_{(0,1)})},\qquad
\Psi_2^{(1\pm1,0)} \simeq v_2.
\eeq
Note here that we have taken terms of the order unity.
A leading order contribution to the interaction comes from the kinetic term of $\Psi_1$ which
is of order ${\cal O}(r^{-2})$. On the other hand, the kinetic energy of $\Psi_2$ and
the potential energy contributions start from the order ${\cal O}(r^{-4})$, so we omit them.
The interaction potential is then given by
\beq
U_{(1\pm1,0)}
&=& \pm \frac{v_1^2 \hbar^2}{m_1} \int d^2x\ \vec\nabla \theta_{(1,0)} \cdot \vec\nabla \theta_{(0,1)} \nonumber \\
&=& \pm \frac{(\mu_1 g_2 - \mu_2 g_{12})\hbar^2 \pi}{(g_1g_2-g_{12}^2)m_1} \log\frac{R^2+L^2}{4R^2}, 
\label{potform2}
\eeq
where $L$ is an infrared cut-off parameter; see Appendix for the details. 
Unlike the case of the leading term in the potential (\ref{eq:pot}) 
between $(1,0)$ and $(0,1)$ vortices, the potential $U_{(1\pm1,0)}$ 
depends on the chemical potential. 
We also note that it depends on the infrared cutoff $L$ but not on 
the ultraviolet cutoff $\xi$.

The inter-vortex force 
$F_{(1\pm1,0)} =  - {d U_{(1\pm1,0)} \over 2 d R}$ is then 
\beq
F_{(1\pm1,0)} 
&=& \pm \frac{(\mu_1 g_2 - \mu_2 g_{12})\hbar^2 \pi}{(g_1g_2-g_{12}^2)m_1} 
\left(\1{R} - {R \over R^2 + L^2}\right) \nonumber \\
&\to& 
\pm \frac{(\mu_1 g_2 - \mu_2 g_{12})\hbar^2 \pi}{(g_1g_2-g_{12}^2)m_1} 
\1{R},
\label{eq:(2,0)force}
\eeq
where $\to$ denotes the large volume limit $L \to \infty$.
This $1/R$ force is well known for 
vortices in the scalar BEC, scalar superfluids and the XY model, 
and global vortices in relativistic field theories \cite{Pethickbook,Shellard:1987bv}.
The correction term for finite volume $L$ 
can be found in the second term in the brace in the 
first line. 
This term might not be so familiar but has been obtained previously 
in \cite{Nakano:2007dq} for global non-Abelian vortices 
in QCD. 

In the same way, 
the interaction potential between $(0,1)$- and $(0,\pm 1)$-vortices are given by
\beq
U_{(0,1\pm1)}(R)
= \pm \frac{(\mu_2 g_1 - \mu_1 g_{12})\hbar^2 \pi}{(g_1g_2-g_{12}^2)m_2} \log\frac{R^2+L^2}{4R^2}.
\eeq

\section{Numerical Analysis}

\begin{figure*}
\begin{center}
\includegraphics[width=15cm]{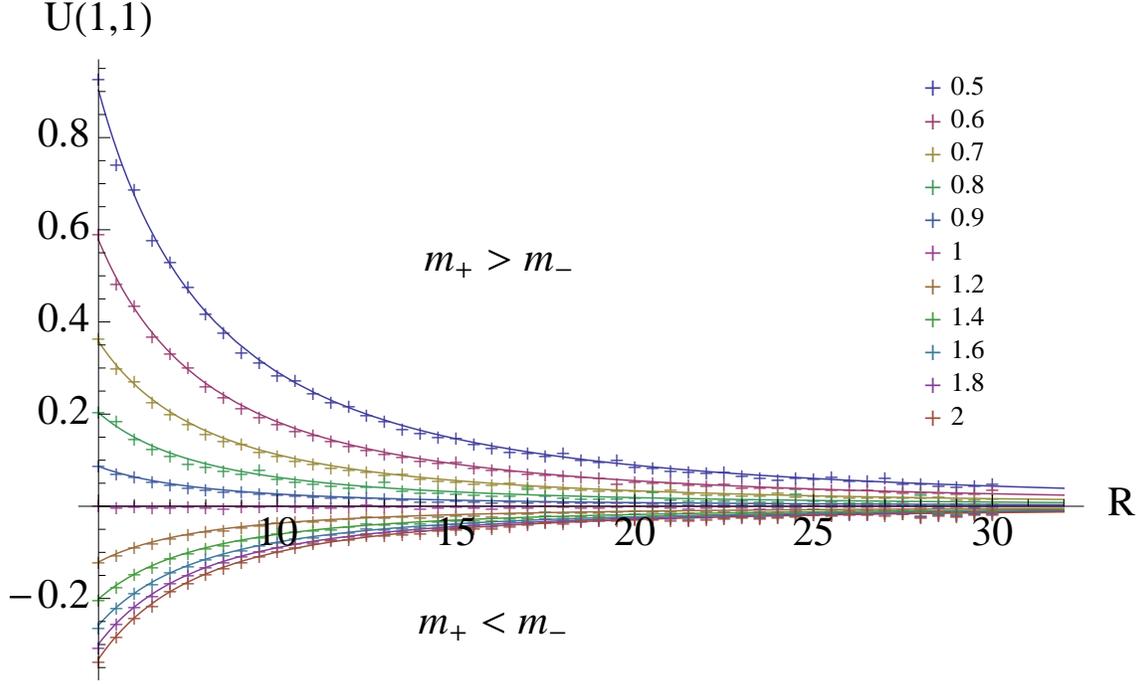}
\caption{The inter-vortex potential $U_{(1,1)}(R)$ 
for $m_-=(0.5,0.6,0.7,0.8,0.9,1,1.2,1.4,1.6,1.8,2)$ with
$m_+ = 1$. Solid lines are asymptotic inter-vortex forces (Abrikosov ansatz) 
which are analytically obtained in Eq.~(\ref{eq:pot}).
}
\label{fig:pot2}
\end{center}
\end{figure*}

Let us numerically verify the interaction potential analytically obtained 
in Eqs.~(\ref{eq:pot}). 
For simplicity, we consider a special case $m_1 = m_2 = m$, $g_1=g_2=g$ and $\mu_1=\mu_2=\mu$.
Then the asymptotic behaviors of the profile functions in Eqs.~(\ref{eq:gp_asym1}) and (\ref{eq:gp_asym2}) are
rewritten as follows,
\beq
f_i + h_i &=& 2 - \frac{1}{m_+^2 r^2} + {\cal O}(r^{-4}), \hspace{3mm}
\label{eq:asym_sp1}\\
f_i - h_i &=& - \frac{1}{m_-^2 r^2} + {\cal O}(r^{-4}), \hspace{3mm}
\label{eq:asym_sp2}
\eeq
for $i={(1,0)}$ and ${(0,1)}$, 
where the mass parameters $m_+$ and $m_-$ are defined by
\beq
m_+^2 \equiv \frac{4m\mu}{\hbar^2}, \quad 
m_-^2 \equiv \frac{4m\mu}{\hbar^2} \frac{g-g_{12}}{g+g_{12}}.
\eeq
The inverse numbers of $m_+$ and $m_-$ give the healing lengths 
associated with 
the mass component $f_{i}+h_{i}$ and the spin component $f_{i}-h_{i}$, respectively.
The inter-vortex potential Eq. (\ref{eq:pot}) is then expressed as
\beq
U_{(1,1)}(R) = \frac{\pi}{2} v^2 \frac{m_{+}^2 - m_{-}^2}{m_{+}^2 m_{-}^2} \frac{1}{R^2} \log \frac{R}{\xi}, 
\label{eq:analitic_pot}
\eeq
where $v^2 = \hbar^2 \mu/ m (g+g_{12})$ is defined in Eq.~(\ref{eq:v}).

To obtain the inter-vortex potential numerically,
we use a sort of the imaginary time propagation of the GP equation as
\beq
\left( -\frac{\hbar^2}{2m}\nabla^2 - \mu + g |\Psi_1({\bf x})|^2 + g_{12}|\Psi_2({\bf x})|^2\right)\Psi_1({\bf x},\tau) \nonumber \\
= -D_1({\bf x})\p_\tau\Psi_1({\bf x},\tau),
\\
\left( -\frac{\hbar^2}{2m}\nabla^2 - \mu + g |\Psi_2({\bf x})|^2 + g_{12}|\Psi_1({\bf x})|^2\right)\Psi_2({\bf x},\tau) \nonumber \\
= -D_2({\bf x})\p_\tau\Psi_2({\bf x},\tau),
\eeq
where $\tau$ is the imaginary time and $D_1$ and $D_2$ are positive coefficients.
While $D_i$ is set to be a constant in the usual imaginary time propagation, 
we consider the coefficient $D_i=D_i({\bf x})$ with space-coordinate dependence. 
An advantage of using $D_i({\bf x})$ is that one can effectively fix the position of vortex 
(the zeros of $\Psi_i$) during the numerical calculation, if one 
choose $D_i({\bf x})$ appropriately. In order to attain this, we choose 
a function $D_i({\bf x}) = A \nabla^2 \log \left(|{\bf x}- {\bf a}_i|^2 + \epsilon^2\right) + c$ 
where ${\bf a}_i$ stands for the $i$-th vortex position and $A$ and
$c$ are positive constants. 
The value of $A$ is taken as an extremely large value to fix the profile of 
the wave function only near the vortex cores. Also, $\epsilon$ should be sufficiently small.
We chose $A=80000$, $\epsilon = 0.01$ and $c=0.1$ in our numerical computation.
We take the Abrikosov ansatz given in Eqs.~(\ref{eq:abrikosov1}) and (\ref{eq:abrikosov2}) as the
initial condition at $\tau=0$ and minimize the energy under the imaginary time evolution. 
After the solutions converge sufficiently, we calculate the 
interaction energy Eq. (\ref{eq:potentia}). 

Throughout our numerical computation below, we will set $m/\hbar^2 = 1$ and $v^2 = 1$.
Then we regard $m_+$ and $m_-$ as independent parameters of the GP equations and perform
the numerical calculation by varying them.
Remember that $m_+ < m_-$ corresponds to $g_{12}<0$ (attractive force) 
whereas $m_+ > m_-$ corresponds to $g_{12}>0$ (repulsive force).
No net interaction exists accidentally when $m_+ = m_-$.

\begin{figure}[ht]
\begin{center}
\includegraphics[width=7cm]{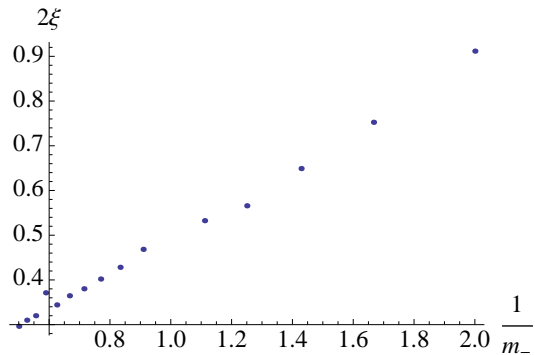} 
\caption{
The dependence of the healing length $2\xi$ on 
$1/m_-$ with fixed $m_+=1$. 
}
\label{fig:healing}
\end{center}
\end{figure}

The result is shown in Fig.~\ref{fig:pot2}.
We compare the inter-vortex potential obtained numerically 
and the one obtained analytically.
As can be seen, the analytic results reproduce the numerical results quite well.
We have only one fitting parameter $2\xi$ which is the 
the short-range cut-off.
The values $2\xi$ for various choice of 
$m_-$ for fixed $m_+=1$ 
are shown in Fig.~\ref{fig:healing}.
We find the linear dependence of 
the short-range cut-off $\xi$ on the healing length $1/m_-$.

\section{Summary and Discussion}
We have studied the asymptotic interaction 
between half-quantized vortices, i.e.,  $(1,0)$ and $(0,\pm 1)$ vortices 
winding around $\Psi_1$ and $\Psi_2$, respectively 
in the two-component BEC. 
Since the two components interact only through the density,  
the $(1,0)$-vortex does not directly experience 
the circulation of the $(0,\pm 1)$-vortex 
so that the result does not depend on the signature 
of the winding number.
The leading order of the force between them is found to be 
$\sim [\log (R/\xi) - 1/2 ]/R^3$ 
in contrast to the one between the same kind of vortices $\sim 1/R$, 
which is also well known as the force between vortices in 
scalar BEC, scalar superfluid and the XY model, and 
global vortices in relativistic field theories.
We  have first derived it analytically using 
the Abrikosov ansatz and the asymptotic profile functions 
of $(1,0)$- and $(0,1)$-vortices. 
We have then confirmed it numerically 
with using the extended imaginary time method for the GP equations.
We have found 
that the short-range cut-off parameter $\xi$ 
of the vortex interaction 
linearly depends on the healing length $1/m_-$.

Our results suggest a bound state of 
$(1,0)$ and $(0,\pm 1)$ vortices for $g_{12}<0$.
While a set of $(1,0)$ and $(0,1)$ 
is expected to form a stable integer (mass) vortex $(1,1)$, 
it is a nontrivial question if $(1,0)$ and $(0,-1)$ vortices 
form a bound state, which should be called a (pseudo-) spin vortex.
Also, one expects no stable bound states for $g_{12}>0$. 
Although there should be instabilities for large separation at least, 
it does not exclude a possibility of 
a metastable bound state at short distance. 
To address these questions, 
we need to know a short range interaction or stability analysis 
of the bound states, 
which remains as a future problem.

Multiple vortices will constitute a vortex lattice 
in experiments of multicomponent BECs under the rotation 
\cite{Schweikhard}. 
Ample phase diagram of the vortex lattices 
was predicted in \cite{Muller} 
and was numerically obtained in \cite{Kasamatsu1,Kasamatsu2}. 
Since we have obtained the analytic expression of the inter-vortex 
forces, we expect to 
explain the vortex phase diagram analytically. 
Especially, we expect that the difference of the force 
$[\log (R/\xi) - 1/2]/R^3$ between $(1,0)$ and $(0,1)$ vortices and 
the one $1/R$ between the same kind of vortices 
will determine it.  

Our method should be extended to spinor BECs, which 
remains as an interesting future problem. 
On the other hand multicomponent systems in relativistic field theories 
are common in QCD 
such as the linear sigma model for the chiral phase transition 
and the Landau-Ginzburg model for 
color superconductors at high baryon density \cite{Balachandran:2005ev}.
In these models, order parameters are matrices as in superfluid $^3$He rather than vectors,  
and consequently there exist 
non-Abelian vortices \cite{footnote2}: non-Abelian global vortices in the chiral phase transition 
\cite{Balachandran:2002je} and 
non-Abelian semi-superfluid vortices in 
color superconductors at high baryon density \cite{Balachandran:2005ev}. 
Inter-vortex forces have been calculated at leading order 
for non-Abelian global vortices \cite{Nakano:2007dq}
and non-Abelian semi-superfluid vortices
\cite{Nakano:2007dr}, see \cite{Nakano:2008dc} for a review. 
Calculation in the present paper will give the next leading 
order $[\log (R/\xi) - 1/2]/R^3$ to them. 
Especially the force $\cos \alpha/R$ between non-Abelian global vortices 
 at the leading order vanishes for a particular choice 
($\alpha = \pm \pi$) of internal orientations of 
vortices \cite{Nakano:2007dq}, and therefore 
the next leading order term proportional to $[\log (R/\xi) - 1/2]/R^3$ 
becomes a dominant contribution. 
An extension of our results to these cases should be important to consider 
a possibility of vortex lattice phases in heavy-ion collisions 
or in a neutron star core, as in two-component BEC 
\cite{Muller,Kasamatsu1,Kasamatsu2}. 

\begin{acknowledgments}
The work of M.E. is supported by Special Postdoctoral Researchers
Program at RIKEN.
The work of K.K. and M.N. is supported in part 
by Grant-in-Aid for Scientific Research 
(Grant No.~21740267 (K.K.) and No.~20740141(M.N.)) 
from MEXT, Japan.
\end{acknowledgments}

\appendix

\section{Derivation of Eqs. (\ref{eq:pot}) and (\ref{potform2})}

\begin{figure}
\begin{center}
\includegraphics[width=8cm]{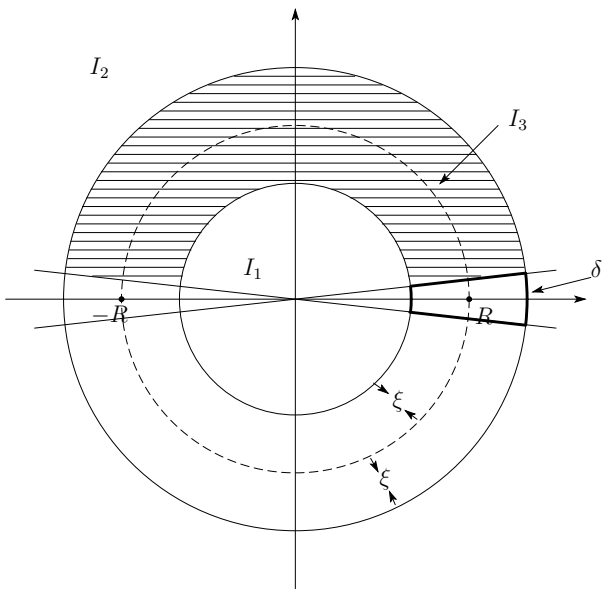}
\caption{The integral region to calculate Eq. (\ref{startintegral})}
\label{integralregion}
\end{center}
\end{figure}
In Eq. (\ref{eq:pot}), we have to evaluate the integral 
\beq
I &=& \int d^2 x \frac{1}{r_{{(1,0)}}^2 r_{{(0,1)}}^2} =  \int d^2 x \frac{1}{A_0},\\
A_0 &\equiv& r^4 + R^4 - 2 r^2 R^2 \cos 2 \theta. 
\label{startintegral}
\eeq
To this end we will use a formula 
\beq
\int^{2\pi}_0d\theta \frac{1}{A+B\cos2\theta} = \frac{2\pi}{\sqrt{A^2-B^2}},
\label{eq:koushiki}
\eeq
for $A > |B|$.
To evaluate Eq.~(\ref{startintegral}), we divide the integral region 
as shown in Fig.~\ref{integralregion}. 
In addition to $I_{1}$ and $I_{2}$, we take into account the contributions $I_{3}$ 
from the strip of width $2\xi$. Since the integrand diverges at $(x,y) = (\pm R,0)$,
we introduce an ultraviolet cut-off $\xi$. Then, we will remove the small regions 
that includes the points of vortex positions. 
Hence, the total integral is written as
\begin{equation}
I_{\rm cut\, off} = I_{1} +  I_{2} + 2 I_{3}.
\end{equation} 
For $R \gg \xi$ the integral $I_{1}$ and $I_{2}$ is calculated as 
\begin{eqnarray}
I_{1} \!\!&=& \!\! \int_{0}^{R-\xi} \frac{2 \pi r dr }{R^{4} - r^{4}} \simeq \frac{\pi}{2 R^{2}} 
\left( \log \frac{R}{\xi} + {\cal O}\left(\frac{\xi}{R}\right)\right), \\
I_{2}  \!\!&=& \!\! \int_{R + \xi}^{\infty} \frac{2 \pi r dr}{R^{4} - r^{4}} \simeq \frac{\pi}{2 R^{2}} 
\left( \log \frac{R}{\xi} + {\cal O}\left(\frac{\xi}{R}\right)\right),
\end{eqnarray}
where we have used Eq.~(\ref{eq:koushiki}).
The remaining integral 
\begin{equation}
I_{3} = \int_{R-\xi}^{R+\xi} dr \int_{\delta}^{\pi-\delta} d\theta \frac{r}{A_0}
\end{equation}
with $\xi/R \ll 1$ and $\delta \ll 1$ is evaluated as follows. 
Note that $\cos 2\theta \le \cos 2\delta = 1 - 2\delta^2 + \cdots < 1 - \delta^2$ and
$A_0 > (r^2-R^2)^2 + 2r^2 R^2 \delta^2 \ge 2r^2 R^2 \delta^2$.
Thus, we have the following inequality
\beq
0 \le I_3 \le \frac{\pi-2\delta}{2R^2\delta^2}\log \frac{R+\xi}{R-\xi} \simeq \frac{\pi}{R^2\delta^2}\frac{\xi}{R}.
\eeq
Thus, for any $\delta$, one can choose sufficiently small $\xi$, so that
$I_3$ becomes negligibly small. 

In summary, we get 
\begin{equation}
I_{\rm cut-off} = \frac{\pi}{R^2} \left(\log \frac{R}{\xi} 
+ {\cal O}\left(\frac{\xi}{R} \right)\right).
\end{equation}

Next, we calculate the integration in Eq.~(\ref{potform2}),
\beq
J &=& \int d^2x\ \vec\nabla \theta_{(1,0)} \cdot \vec\nabla\theta_{(0,1)}
= \int d^2x\ \frac{r^2-R^2}{r_{(1,0)}^2r_{(0,1)}^2} \non
&=& \int drd\theta \frac{r(r^2-R^2)}{r^4+R^4-2r^2R^2\cos2\theta}.
\eeq
By using Eq.~(\ref{eq:koushiki}), one can first perform the integration in $\theta$
and then integrate with respect $r$ as
\beq
J &=& \int_0^\infty dr \frac{2\pi r(r^2-R^2)}{\sqrt{(r^4-R^4)^2}} \non
&=& \lim_{L\to\infty} \left[-\int_0^R \!\! \frac{2\pi rdr}{r^2+R^2} 
+ \int_R^L \frac{2\pi rdr}{r^2+R^2} \right] \non
&=& \lim_{L\to\infty}
\pi \log \frac{L^2+R^2}{4R^2}.
\eeq


\end{document}